\documentclass[12pt]{article}

\usepackage{amsmath}
\usepackage{amssymb}
\usepackage{verbatim}

\makeatletter

   \@addtoreset{equation}{section}
\makeatother

\newcommand{\be}{\begin{equation}}
\newcommand{\ee}{\end{equation}}
\newcommand{\bea}{\begin{eqnarray}}
\newcommand{\eea}{\end{eqnarray}}

\begin{document}

\vskip 12mm

\begin{center}

{\Large \bf Solutions for Mixed States in Open Bosonic String Theory}
\vskip 10mm
{ \large  Dimitri Polyakov$^{a,b,}$\footnote{email:polyakov@scu.edu.cn;polyakov@sogang.ac.kr}
}

\vskip 8mm
$^{a}$ {\it  Center for Theoretical Physics, College of Physical Science and Technology}\\
{\it  Sichuan University, Chengdu 6100064, China}\\
\vskip 2mm

$^{b}$ {\it Institute of Information Transmission Problems (IITP)}\\
{\it  Bolshoi Karetny per. 19/1, Moscow 127994, Russia}

\end{center}

\vskip 15mm

\begin{abstract}

We describe the family of normalizable solutions
in linearized open string field theory, defined by
$Q\Psi_0=0$ ($Q$ is BRST charge) understood in the sense
$<<Q\Psi_0,\Phi>>=0$ for an arbitrary string field $\Phi$.
The solutions depend on shifted partition numbers and are parametrized
in terms of values of $\zeta$-function at pairs of positive numbers greater than 2.
We argue that the operators, defined by these solutions, create
mixed quantum-mechanical states by acting on the vacuum (as opposed to standard vertex operators,
creating the pure states with definite masses and spins).

\end{abstract}
%\vfill
%\end{titlepage}

\vskip 12mm

\setcounter{footnote}{0}

%===================
\section{\bf Introduction}
%====================
In $D$-dimensional open bosonic string theory \cite{polya} the action in conformal gauge is given by
\bea
S\sim{\int{d^2z}\lbrace{\partial{X_m}\bar\partial{X^m}+b{\bar\partial}c+{\bar{b}}\partial{\bar{c}}\rbrace}+S_{Liouville}}
\nonumber \\
m=0,...,D-1
\eea
and the nilpotent BRST operator $Q^2=0$ \cite{brst} is given by
\bea
Q=\oint{{dz}\over{2i\pi}}
\lbrace{cT-bc\partial{c}}\rbrace
\equiv
\oint{{dz}\over{2i\pi}}
\lbrace{-{1\over2}{c}\partial{X_m}\partial{X^m}+bc\partial{c}+...}
\rbrace
\eea
where $T$ is the full stress-energy tensor and we skipped the Liouville terms
in the second integral (as they will play no role in the rest of the paper).
The physical spectrum of string theory, modulo BRST-exact states,  is defined by vertex operators $\lbrace{V}\rbrace$
satisfying
\be
QV=0
\ee
(this equation involves anticommutators or commutators, for unintegrated and integrated pictures
of the vertices respectively).
The equation (1.3), defining the physical spectrum of bosonic string: ${\lbrace}|\Psi>\rbrace{=}{\lbrace}V|0>\rbrace$ 
can  also be viewed as a linearized
limit of open string field theory equation of motion \cite{witf, wits} (e.g. $Q\Psi+\Psi\star\Psi=0$ for the cubic OSFT).
The equation (1.3) has two well-known classes of solutions: the local operators -  dimension 0  primary
fields of ghost number $+1$:
\be
V=cP(\partial{X},\partial^2{X},...)e^{ipX}\varphi(p)
\ee
and also the worldsheet integrals of dimension 1 primaries with ghost number zero:

\be
V=\varphi(p){\int{{dz}P(\partial{X},\partial^2{X},...)e^{ipX}}}
\ee
where  $P$ are polynomials in the derivatives of $X$ defining the masses of excitations
(for simplicity, we limit ourselves to the open string case only).

These solutions are related by the $b-c$ picture-changing transformation, defined
by the BRST-invariant $b-c$ picture-changing operator $Z=:b\delta(T):$ \cite{selfff}.
From the point of view of the second-quantized theory, (1.3) as a classical equation,
just like Klein-Gordon equation (a field-theoretic analogue of (1.3)) is a classical equation for a classical field
(infinite number of oscillators). At the same time, it is well-known that the Klein-Gordon equation also can be thought as a
an equation defining a $quantum-mechanical$ wavefunction for a single particle (oscillator), at least as long as no interaction
terms are added. In the similar spirit, the solutions (1.4), (1.5) of (1.3), 
acting on the vacuum in open string theory,  can be understood as wavefunctions of particular open string excitations. 
From the quantum-mechanical point of view, (1.4)  and (1.5) define the pure states, characterized by definite
masses and spins (eigenvalues of Casimir operators of Poincare algebra in space-time).
The solutions (1.4), (1.5) defining the physical spectrum of an open string, are the only on-shell solution of (1.3).
In this letter we show that, apart from (1.4) and (1.5), there exists another class of solutions of (1.3) in open string field theory,
formulated off-shell. These solutions are given by operators expressed  in terms of infinite formal series in derivatives of $X$
and are understood in terms of the vanishing of OSFT correlators : 
\be
<<Q\Phi;\Psi>>=0
\ee
where $\Psi$ is any string field and $\Phi$ is the solution that we aim to describe.
The completeness of the operator algebra in CFT then ensures that is the vanishing occurs for a two-point
function for any $\Psi$, the same would be true for the insertion of $Q\Phi$ into any SFT correlator
with any number of points. The  condition (1.6) is far stronger then, e.g., BRST-triviality, since in string field
theory BRST-exact insertions generally do not lead to the vanishing of the correlators, as the operators are off-shell.

Being  infinite formal series in derivatives of $X$, the solutions that we describe, mix the modes with 
different masses and spins. As such, acting on a vacuum, they cannot define any pure state with a wavefunction (unlike  
the solutions (1.4), (1.5)), but instead constitute the summation over the ensemble of operators with different spins and masses.
In this sense, these solutions, acting on the vacuum, appear to correspond to mixed quantum-mechanical states in open string theory.
The coefficients in the formal series defining the solutions are then related to the eigenvalues of the density matrices
describing the states created by (1.6).
It turns out that these coefficients are expressed in terms of shifted partition numbers of the spin values appearing in the series,
and can be labelled by  values of the Riemann's zeta-function at certain positive parameters.
Below we shall describe these solutions in details.

\section{\bf BRST-invariant Mixed States}

Consider  a quantum-mechanical subsystem $A$ being a part of a larger system $B$.
In general, a quantum-mechanical state  of $A$ cannot be described by a single wavefunction $\psi_A$
depending only on the particulars of the smaller subsystem. 
Instead, it is described by a density matrix reflecting the interaction between $A$ and $B$,
leading to the entanglement of $A$ with $B$ \cite{eins, raams}
In particular, such a system can be described by the $ensemble$ of states $\psi^i_A$ ($i=1,...,N$)
($N$ can be finite or infinite; $N=1$ corresponds to the pure state)
This ensemble effectively diagonalizes the density matrix, which is given by:
\be
\rho_A=\sum_i\gamma_i|\psi_A^i><\psi_A^i|
\ee
where $\gamma^i$ are the $classical$ probabilities reflecting the quantum uncertainty of the subsystem.
The solutions of the equation (1.3) that we we shall present below are given by the infinite formal series
that we shall interpret as summations over ensembles of states with different masses and spins,
with the coefficients corresponding to the eigenvalues of the density matrix, defined by this solution.

 For simplicity, let us start from the $D=1$ case.
which will be straightforward to generalize to higher space-time dimensions.
The solution that we are looking for, particularly satisfies Siegel gauge condition:
\begin{equation}
b_0\Psi=0
\end{equation}
with the ghost number 1 and with the following expansion in infinite formal series
in derivatives of $X$:
\begin{eqnarray}
\Psi_0=c\sum_{N=1}^\infty\sum_{p=1}^N\sum_{N|n_1...n_p}
\alpha_{n_1...n_p}{{\partial^{n_1}X}\over{n_1!}}...{{\partial^{n_p}X}\over{n_p!}}
\end{eqnarray}
where $\sum_{N|n_1...n_p}$ stands for the summation over ordered length $p$ partitions of N:
\begin{eqnarray}
N=n_1+...+n_p
\nonumber \\
n_1\geq{n_2}...\geq{n_p}>0
\end{eqnarray}
and $\alpha_{n_1...m_p}$ are some coefficients.
The numbers $N$ and $p$ are thus useful parameters of such an expansion;
although not directly related to higher-spin currents in space-time in $D=1$, in higher space-time
dimensions
 conformal dimension (worldsheet spin) $N$ of a string field component 
actually can be related 
to the space-time spin $N$ of the component,
with the contributions from different $p$ looking like ``Stueckelberg-like'' terms \cite{vas, sagn, stue, selfbook}.
It is therefore convenient to cast $\Psi_0$ as
\begin{eqnarray}
\Psi_0\equiv{c}\sum_{N=1}^\infty\sum_{p=1}^{N}\Psi_0^{(N;p)}
\end{eqnarray}
(with $\Psi_0^{(N;p)}$ read off directly from the previous equation)
Our initial goal is to find the choice of the coefficients $\alpha$ for which
 $\Psi_0$  satisfies:
\begin{equation}
<<Q\Psi_0,\Psi>>\equiv<Q\Psi_0(0)I\circ\Psi(0)>=0
\end{equation}
for {any} string field $\Psi$ ( {\bf  not} necessarily in the Siegel's gauge) 
Since $\Psi$ is arbitrary, this identity, once true for any two-point correlator,
will also be true for the insertion of $Q\Psi_0$ into any other SFT correlator,
due to the completeness
 of the full operator algebra in CFT, which is equivalent
to the statement that $Q\Psi_0$ vanishes identically.
Here the double brackets stand for the standard OSFT correlator and
the conformal transformation $I(z)=-{1\over{z}}$
maps $\Psi$ to infinity.

Let us start with evaluating $Q\Psi_0$.
Simple calculation gives:
\begin{eqnarray}
Q(c\sum_{N,p}\Psi_0^{N;p})=\sum_{N,p}(N-1)\partial{c}c
\Psi_0^{N;p}
\nonumber \\
+\sum_{N,p}\sum_{N|n_1...n_p}\sum_{j=1}^p\sum_{k=2}^{n_j}{{\partial^{k}cc}\over{k!}}{{\alpha_{n_1...n_p}\partial^{n_1}X
...\partial^{n_{j-1}}X\partial^{n_j-k}X\partial^{n_{j+1}}X...{\partial{n_p}}X}\over{n_1!...n_{j-1}!(n_j-k)!n_{j+1}!...n_p!}}
\nonumber \\
+\sum_{N,p}
\sum_{N|n_1...n_p}
\sum_{1\leq{i}<j\leq{p}}{{\partial^{n_i+n_j+1}cc}\over{(n_i+n_j+1)!}}
\nonumber \\
\times
{{\alpha_{n_1...n_p}\partial^{n_1}X...
\partial^{n_{i-1}}X\partial^{n_{i+1}}X...\partial^{n_{j-1}}X\partial^{n_{j+1}}X...\partial^{n_p}X}\over
{n_1!...n_{j-1}!n_{i+1}!....n_{j-1}!n_{j+1}!...n_p!}}
\end{eqnarray}

It can be shown, however that,  with $\Psi_0$ ansatz given by (2.3)  only the first term
in $Q\Psi_0$ (2.7), proportional to $\partial{c}c$, contributes
to the correlator $<<Q\Psi_0\Psi>>$ for any $\Psi$.
To prove this,
it is convenient to bosonize the $b-c$ ghosts
according to:
\bea
b=e^{-\sigma}
\nonumber \\
c=e^\sigma
\eea

Note that  $Q\Psi_0$ has ghost number 2, so  the only $\Psi$  components
contributing to the correlator are those having ghost number 1.
The operators having ghost number 1 in general have the ghost part proportional to 
$$:\sim\partial^{m_1}b...\partial^{m_r}b\partial^{n_1}c...\partial^{n_{r+1}}c:\sim{:G(\partial\sigma,\partial^2\sigma,...):}e^\sigma$$
where $m_j,n_j$ are non-negative integers and $G$ is some polynomial in derivatives of $\sigma$. 
First of all, it is clear that only the terms with $r=0$ or $1$ can contribute
(otherwise there would be $b$-fields left with no contractions). Let us first check our claim for $r=0,n_1=0$ and then generalize 
it to the arbitrary case. 
 In the case of $r=0,n_1=0$ ($\Psi$-field proportional to the $c$-ghost)
 the ghost part of the correlator
 $<<Q\Psi_0\Psi>>$ has the form $<\partial^k{c}c(0)I{\circ}c(0)>$.
It is then easy to check that the only nonzero correlator is the one 
with $k=1$. Indeed, write 
$I\circ{c}=({{dI}\over{dz}})^{-1}|_{z=0}c(\infty)={\lim_{w\rightarrow{\infty}}}w^{-2}c(w)$. 
Then, at  $k=1$,

\begin{eqnarray}
<\partial{c}c(0)I{\circ}c(0)>={\lim_{w\rightarrow\infty}}w^{-2}<\partial{c}c(0)c(w)>
={\lim_{w\rightarrow{\infty}}}w^{-2}w^2=1
\end{eqnarray}
At $k=2$,
\begin{equation}
<\partial^2{c}c(0)I{\circ}c(0)>={\lim_{w\rightarrow\infty}}w^{-2}<\partial^2{c}c(0)c(w)>
={\lim_{w\rightarrow{\infty}}}w^{-2}(-{2{w}})=0
\end{equation}
For higher $k>2$ the ghost correlators also vanish identically; that is, using 
the bosonization (2.8) we write 
\bea
\partial^kc=B^{(k)}(\sigma(z);z)
\eea
 where 
$B^{(k)}(\sigma(z);z)=B^{(k)}(\partial\sigma,...\partial^k\sigma)$ is the degree $k$ Bell polynomial in derivatives of $\sigma$
(see (2.16) for the precise definition).
 Its OPE
with $e^\sigma$ has the form: 
\begin{equation}
B^{(k)}(\sigma(z);z)(z)e^\sigma(w)=(z-w)^{-1}
k:B^{(k-1)}(\sigma(z);z)(z)e^\sigma:(w)+O(z-w)^0,
\end{equation} 
so 
\begin{equation}
:\partial^kcc:={k}:B^{(k-1)}(\sigma(z);z){e^{2\sigma}}:
\nonumber
\end{equation}
As it is clear from the OPE (2.12), for $k>2$ the polynomial 
$:B^{(k-1)}(\sigma(z);z):(0)$ cannot fully contract with the $c$-ghost at infinity
and all such correlators vanish identically.
This constitutes the proof that only the terms proportional to
$N\partial{c}c\Psi_0^{(N,p)}$ ($k=1$)  in $Q\Psi_0$ contribute to the correlator
$<<Q\Psi_0\Psi>>$ with the components of $\Psi$ satisfying $r=0,n_1=0$.
Now let us show that, once this is true  for $r=0,n_1=0$, this is also true for
arbitrary components of $\Psi$. For the reasons pointed out above, it is sufficient to show  that this is
 the case for $r=1$,i.e. 
for the components of $\Psi$ with the ghost structure $\sim:\partial^{m_1}b\partial^{n_1}c\partial^{n_2}c:$
First of all, note that, since the correlator $<<\partial^k{c}c(0) I\circ{c}(0)>>=0$  on the half-plane for $k>1$, 
it also vanishes under
any conformal transformation: $z\rightarrow{f(z)}$ of the half-plane. Now let us consider the half-plane correlator
$<<\partial^{k}c{c}(0)(I\circ({\partial^{m_1}b\partial^{n_1}c\partial^{n_2}c})(w\rightarrow\infty)>>$ 
(for the certainty, on the upper half-plane) and apply the conformal transformation ${z}{\rightarrow}f(z)=e^{iz}$.
This transformation is well-defined everywhere on the upper half-plane (including the real axis) and vanishes 
exponentially fast at infinity. Under this transformation,
the $:\partial^{m_1}b\partial^{n_1}c\partial^{n_2}c(z):$ 
operators transform as
\begin{eqnarray}
:\partial^{m_1}b\partial^{n_1}c\partial^{n_2}c:(w\rightarrow\infty)\equiv{:H(\partial\sigma,\partial^2\sigma,...)e^\sigma:}(w\rightarrow\infty)
\nonumber \\
\rightarrow
{\lim_{w\rightarrow\infty}}{\lbrace}S(m_1|n_1,n_2)(e^{iw};w)c(w)+O(e^{iw})\rbrace
\end{eqnarray}
where $H$ is some polynomial in derivatives of $\sigma$ and
we skipped the terms of orders of $e^{iw}$ and higher (suppressed exponentially when $w$ is taken to infinity).
Next, $S(m_1|n_1,n_2)(e^{iw};w)$ are the combinations of 
generalized Schwarzians of the conformal transformation $z\rightarrow{e^{iz}}$
of the upper half-plane, appearing as a result of the regularization of the internal singularities in 
operator products between the derivatives of the $b$ and $c$-ghosts. For the exponential conformal transformation
of the half-plane $S(m_1|n_1,n_2)(e^{iw};w)$ are constant numbers that do not depend on $w$ (see below for the 
discussion of some essential properties of the generalized Schwarzians).
For this reason, the correlators $\lim_{w\rightarrow\infty}<\partial^k{c}c(0)\partial^{m_1}b\partial^{n_1}c\partial^{n_2}c(w)>$,
computed on the Riemann surface as the  result of the conformal transformation of the upper half-plane,
are proportional to the correlators $\lim_{w\rightarrow\infty}<\partial^k{c}c(0)c(w)>$ on the same Riemann surface
(with the coefficients given by  constant generalized Schwarzian factor) and  therefore vanish  for $k>1$.
This constitutes the proof that only the terms proportional to
$N\partial{c}c\Psi_0^{(N,p)}$ need to be considered in $Q\Psi_0$, if $\Psi_0$ has the form (2.3).
We are now prepared to analyze the correlator
$<Q\Psi_0(0)I\circ\Psi(0)>$ for 
$\Psi_0$ of the form (2.3) and 
an arbitrary string field $\Psi$. 

The string fields of this correlator are located on the halfplane's boundary;
the crucial next step to compute the correlator is again the conformal transformation of the half-plane:
\begin{equation}
z\rightarrow{f(z)}=e^{iz}
\end{equation}
taking the upper half-plane to compact Riemann surface,
with $Q\Psi_0$ taken from zero to 1 and $\Psi$ from infinity to zero.
This conformal transformation (which we will also refer to as the ``singularization transformation''
) maps the  upper half-plane to a compact Riemann surface which we shall call the ``singularoid''.
 
Consider the behavior of the $<<Q\Psi_0,\Psi>>$ correlator under such a conformal map.
For that, one crucial relation that we shall need is
the transformation law of the $:\partial^{n_1}X\partial^{n_2}X:(z)$-operator
under $z\rightarrow{f(z)}$, given by
\begin{eqnarray}
{1\over{n_1!n_2!}}:\partial^{n_1}{X}\partial^{n_2}X:(z)\rightarrow
\nonumber \\
{1\over{n_1!n_2!}}\sum_{k_1=1}^{n_1}\sum_{k_2=1}^{n_2}B_{n_1|k_1}(f(z);z)B_{n_2|k_2}(f(z);z)
:\partial^{k_1}{X}\partial^{k_2}{X}:(f(z))
\nonumber \\
+S_{n_1|n_2}(f(z);z)
\end{eqnarray}
where
$B_{n|k}$ are the incomplete Bell polynomials in the $z$-derivatives of $f$.
The general definition of $B_{n|k}$ is:
\begin{eqnarray}
B_{n|k}(g_1,...g_{n-k+1})=n!\sum{1\over{p_1!...p_{n-k+1}!}}
{({{g_1}\over{1!}})^{p_1}}...
{({{g_{n-k+1}}\over{(n-k+1)!}})^{p_{n-k+1}}}
\end{eqnarray}
with the sum taken over all the non-negative $p_1,...p_{n-k+1}$
satisfying
\begin{eqnarray}
p_1+...+p_{n-k+1}=k
\nonumber \\
p_1+2p_2+...+(n-k+1)p_{n-k+1}=n
\nonumber
\end{eqnarray}
In particular,
the incomplete Bell polynomials $B_{n|k}(f;z)$  
 in  the derivatives 
(or the expansion coefficients) of $f(z)$,
are given by the substitution
$g_k=\partial_z^{k}f(z)\equiv{{d^kf}\over{dz^k}}$
(although the  partial derivative sign is not necessary, we keep it to shorten our notations)
or equivalently

\begin{eqnarray}
B_{n|k}(f(z);z)=n!\sum_{n|n_1...n_k}{{\partial^{n_1}f(z)...\partial^{n_k}f(z)}
\over{n_1!...n_k!q(n_1)!...q(n_k)!}}
\nonumber
\end{eqnarray}
with the sum ${n|n_1...n_k}$ taken 
over all ordered $<n_1\geq{n_2}...\geq{n_k}>0$
length $k$ partitions of $n$ and with $q(n_j)$
denoting the multiplicity of  $n_j$ element of the partition
(e.g.  for the partition $7=2+2+3$ we have $q(2)=2,q(3)=1$,
so the appropriate term would read 
$\sim{{\partial^2{f}\partial^2{f}\partial^3{f}}\over{2!2!3!\times{2!1!}}}$.
Then, $S_{n_1|n_2}(f(z);z)$ are the generalized Schwarzians of the conformal transformation
from the half-plane to the singularoid,
given by \cite{selfnumber}
\begin{eqnarray}
S_{n_1|n_2}(f;z)=
{1\over{n_1!n_2!}}\sum_{k_1=1}^{n_1}\sum_{k_2=1}^{n_2}
\sum_{m_1\geq{0}}\sum_{m_2\geq{0}}\sum_{p\geq{0}}
\sum_{q=1}^p
(-1)^{k_1+m_2+q}2^{-m_1-m_2}(k_1+k_2-1)!
\nonumber \\
\times
{{\partial^{m_1}B_{n_1|k_1}(f(z);z)\partial^{m_2}B_{n_2|k_2}(f(z);z)
B_{p|q}(g_1,...,g_{p-q+1})}
\over{m_1!m_2!p!(f^\prime(z))^{k_1+k_2}}}
\nonumber \\
g_s=2^{-s-1}(1+(-1)^s){{{{d^{s+1}f}\over{dz^{s+1}}}}\over{(s+1)f^\prime(z)}};
s=1,...,p-q+1
\end{eqnarray}
with the sum over the non-negative numbers $m_1,m_2$ and $p$ taken over all the combinations satisfying
$$m_1+m_2+p=k_1+k_2$$
For $n_1=n_2=1$ $S_{1|1}$ becomes the usual Schwarzian derivative (up to the conventional
normalization factor of ${1\over6}$).
Note that the exponential factors proportional to powers of $\sim{e^{iz}}$
cancel out in all the terms of the summation,
so for the conformal transformation that we need, $f(z)=e^{iz}$, the generalized  Schwarzians
$S_{n_1|n_2}$ do not depend on $z$ and are constant.
For the conformal transformation under study, $f(z)=e^{iz}$, the value of the  Bell polynomials $B_{n|p}(f(z);z)$ 
and their derivatives at can be expressed in terms of the Stirling numbers of the second kind $S(n;k)$:
\begin{eqnarray}
B_{n|k}(e^{iz};z)=i^nS(n;k)e^{ikz}
\nonumber \\
\partial^p_z{B_{n|k}}(f(z);z)=i^{n+p}k^pS(n;k)e^{ikz}
\end{eqnarray}
and accordingly, for $f(z)=e^{iz}$ the explicit form of the generalized Schwarzians
can be simplified to give:
\begin{eqnarray}
S_{n_1|n_2}(f;z)=
{1\over{n_1!n_2!}}\sum_{k_1=1}^{n_1}\sum_{k_2=1}^{n_2}
\sum_{m_1\geq{0}}\sum_{m_2\geq{0}}\sum_{p\geq{0}}
\sum_{q=1}^p
(-1)^{k_1+m_2+q}2^{-k_1-k_2}(k_1+k_2-1)!
\nonumber \\
\times
{{i^{-p}S(n_1,k_1)S(n_2,k_2)k_1^{m_1}k_2^{m_2}
B_{p|q}(g_1,...,g_{p-q+1})}
\over{m_1!m_2!p!}}
\nonumber \\
g_s={{2^{-s}cos({{\pi{s}}\over{2}})}\over{s+1}}
\end{eqnarray}
with the summations subject to the same constraints as indicated below (2.17).
The transformation law (2.15) is straightforward to generalize for any monomial
in the derivatives of $X$.
Namely, under $z\rightarrow{f(z)}$ we have

\begin{eqnarray}
:\partial^{n_1}X...\partial^{n_p}X:(z)\rightarrow
\nonumber \\
\sum_{q=1}^{\lbrack{p\over2}\rbrack}
\sum_{\lbrace{1...p}\rbrace\rightarrow{\lbrace}i_1...i_{2q};j_1...j_{p-2q}\rbrace}
\sum_{k_1=1}^{n_{j_1}}...\sum_{k_{p-2q}=1}^{n_{j_{p-2q}}}
S_{n_{i_{1}}|n_{i_{2}}}(f(z);z)...S_{n_{i_{2q-1}}|n_{i_{2q}}}(f;z)
\nonumber \\
B_{n_{j_1}|k_1}(f(z);z)...B_{n_{j_{p-2q}}|k_{p-2q}}(f(z);z):\partial^{k_1}X...
\partial^{k_{p-2q}}X:(f(z))
\end{eqnarray}
where
$\sum_{\lbrace{1...p}\rbrace\rightarrow{\lbrace}i_1...i_{2q};j_1...j_{p-2q}\rbrace}$
stands for the summation over the permutations
$\lbrace{1...p}\rbrace\rightarrow{\lbrace}i_1...i_{2q};j_1...j_{p-2q}\rbrace$
such that $i_1\neq{i_2}...\neq{i_{2q}}\neq{j_1}...\neq{j_{p-2q}}$;
$1\leq{i_k}\leq{p};1\leq{j_k}\leq{p}$ and $i_{2k-1}\leq{i_{2k}}$
(the last constraint is imposed in order to ensure that the redundant
combinations of  
Schwarzians $S_{n_i|n_j}$ do not appear in the permutations).
 
In what follows, we will be particularly interested in the terms
with $p=2q$ in the sum (2.21) that contain no operators but are just the numbers
only depending on $f(z)$. We shall call these terms
{\it pure Schwarzian contributions}, and they will be play an 
important role in the calculations below. To simplify
the notations, it is convenient to write
\begin{eqnarray}
S_{n_1...n_p}(f(z);z)=
\sum_{\lbrace{1...p}\rbrace\rightarrow{\lbrace}i_1...i_{p}\rbrace}
S_{n_{i_{1}}|n_{i_{2}}}(f(z);z)...S_{n_{i_{p-1}}|n_{i_{p}}}(f(z);z)
\end{eqnarray}
with the summation over permutations of $1....p$ defined as above.
We shall call $S_{n_1...n_p}$ the {\it Schwarzian image}
of the operator $\partial^{n_1}X...\partial^{n_p}X$ under the conformal
map $f(z)=e^{iz}$.
We are now prepared to return to the conformal transformation (2.15)
of $<<Q\Psi_0,\Psi>>$.
First, consider the transformation of $I\circ\Psi$ located at infinity.
Because of the  proportionality property of the ghost operators
under the conformal transformation (2.14),  discussed above,
it is sufficient do consider the $\Psi$-operators having the form same as (2.5),
except for possible explicit dependence on the logarithmic $X$-field. 
According to the transformation formula (2.20), each term
in $\Psi$ gets multiplied by $e^{ihz}|_{z\rightarrow\infty}$ with
$h\geq{p-2q};0\leq{2q}\leq{p}$. Therefore all the contributions,
except for the one with $p=2q$ (that is, the one involving the Schwarzian image
$S_{n_1...n_p}$)
 are exponentially dumped and vanish identically
at infinity. So for any positive $N=n_1+...+n_p$ the only surviving 
part in any
component of $\Psi$ upon the conformal transformation (2.14) is the pure 
Schwarzian (which is constant, given by sum of combinations
of the products involving Stirling numbers according to (2.19)).
The only possible exception to it is the component with $N=0$
which, in principle, also may be present in $\Psi$.
This component is just a function of $X$ with no derivatives having the form $:f(X):$. But such a 
component a priori does not contribute to the contractions
with $\Psi_0$ in the correlator (note that $\Psi_0$ by construction contains no $N=0$ terms).
To see this, it is convenient to apply the conformal transformation $I(z)$ to the correlator
$<<Q\Psi_0(0)I{\circ}(:X^n:(0))>>$ for any $n$, taking :$X^n$: from infinity to zero and $\Psi_0$ at 0
to ${\tilde{Q\Psi_0}}$ at infinity, with ${\tilde{Q\Psi_0}}$ having the same form (2.7) as $Q\Psi_0$,
but with some new coefficients ${\tilde{\alpha}}_{n_1...n_p}$, straightforward to determine
from the conformal transformation. Note that $X^n$ doesn't change
as the resulting  conformal transformation applied to it, $I\circ{I}$ , is an identity. 
Then, using the translational invariance, take
$f(X)$ to $z=-\pi$, and apply another transformation $f(z)=e^{iz}$ to the correlator

$$<X^n(-{\pi\over2})(I\circ{{{{Q\Psi}}_0}})(\infty)>.$$ 

Similarly to what we explained before, 
only the pure Schwarzian terms remain out  of ${\tilde{\psi}}_0$ upon the transformation, implying
that the entire correlator is proportional to the pure Schwarzian factor of $X^n$ which does not contract.
But this factor is proportional to 

$$(S_{0|0}(f(z);z))^{n\over{2}}|_{f(z)=e^{iz};z={-{\pi\over2}}}$$
where $S_{0|0}={\log{(f^\prime(z))}}$, i.e. vanishes at $z=-{{\pi\over2}}$.
This shows that the only possible string field component of $I\circ\Psi$,
that does not vanish under $f(z)=e^{iz}$, except for the pure Schwarzian part,
does not contribute to the correlator $<<Q\Psi_0I\circ{\Psi}>>$.
 But then, since only the pure Schwarzian (non-contracting)
terms of $\Psi$ 
contribute to the correlator, the same is true for $Q\Psi_0$; 
therefore we conclude that the correlator $<Q\Psi_0(0)I\circ(\Psi(0))>$
evaluated on the singularoid has the form:

\begin{eqnarray}
<<Q\Psi_0\Psi>>=G_\Psi\sum_{N=1}^\infty\sum_{p=1}^{N}\sum_{N|n_1...n_p}\alpha_{n_1...n_p}S_{n_1...n_p}
\end{eqnarray}
where $G_\psi$ is some constant which only depends
on particulars of $\Psi$ and independent on $\Psi_0$.
The coefficients $\alpha_{n_1...n_p}$ are now to be chosen so that the correlator involving 
the summation over $N$ vanishes.
For that, the first step is to deduce $S_{n_1...n_p}$
(as previously, we consider $p$ even).
Despite the seeming complexity of the Schwarzian image (lengthy sum over combinations of products of individual Schwarzian)
the resulting expression for $S_{n_1...n_p}$ is relatively simple and can be expressed
in terms of shifted partition numbers, depending on $p$ and $N=m_1+....+n_p$. 
To deduce it,
consider the test correlator of $Q\Psi_0$ with ${1\over{p!}}:I\circ(\partial{X})^p:$ 
(multiplied by the c-ghost, as usual) for some $p$.
The relevant terms in the part of  $<<Q\Psi_0,(\partial{X})^p>>$
for a given $N$  are
\begin{eqnarray}
{1\over{p!}}\sum_{N|n_1...n_p}<{{\partial^{n_1}X...\partial^{n_p}X(0)}\over{n_1!..n_p!}}I(\circ(\partial{X})^p(0))>
\nonumber \\
={\lim_{w\rightarrow{\infty}}}
{1\over{p!}}\sum_{N|n_1...n_p}<{{\partial^{n_1}X...\partial^{n_p}X(0)}\over{n_1!...n_p!}}w^{2p}(\partial{X})^p(w))>
(U_{0}(w))
\end{eqnarray}
where $U_0(w)$ is the overlap factor accounting for
for the correlator change  as a result of the integration of conformal Ward identities 
(note that the correlator (2.24), computed naively without this factor would have been proportional to 
to $\sim{w^{p-N}}$, i.e. would have vanished, as the $w^{2p}$-factor due to the conformal transformation
of $(\partial{X})^p$ by $I$ would have been multiplied by $w^{-N-p}$  as a result of the contractions). 

In our case, this factor can be computed explicitly.
Infinitezimally, it is given by the integral
\begin{eqnarray}
{1\over{p!}}\delta_\epsilon
\sum_{N|n_1...n_p}<{{\partial^{n_1}X...\partial^{n_p}X(0)}\over{n_1!..n_p!}}I\circ((\partial{X})^p(0))>
\nonumber \\
=-\lbrack{1\over{2p!}}\oint{{dz}\over{2i\pi}}\epsilon(z)\partial{X}\partial{X}(z);
\sum_{N|n_1...n_p}<{{\partial^{n_1}X...\partial^{n_p}X(\xi)}\over{n_1!...n_p!}}
(\partial{X})^p(w)>\rbrack|_{overlap;\xi=0,w\rightarrow\infty}
\nonumber \\
=
\sum_{N|n_1...n_p}\sum_{j=1}^p{1\over{(p-1)!n_1!...n_{j-1}!n_{j+1}!...n_p!}}
\nonumber \\
\times
<\partial^{n_1}X...\partial^{n_{j-1}}X\partial^{n_{j+1}}X...\partial^{n_p}X(\xi)
(\partial{X(w)})^{p-1}>|_{\xi=0,w\rightarrow\infty}
\nonumber \\
\times
\oint{{dz}\over{2i\pi}}{{\epsilon(z)}\over{(z-\xi)^{n_j+1}(z-w)^2}}
\end{eqnarray}
with one of the $\partial{X}$'s in the stress tensor $T(z)$ acting on the operator at $\xi$
and another on the operator at $w$ (i.e. the infinitezimal overlap transformation
gives the change of the entire correlator under the conformal transformation 
excluding the contributions due to infinitezimal conformal transformations of the vertex operators themselves).
The integral over $z$ is straightforward to evaluate, however, since the conformal transformation by
$I(z)$ only acts on the second operator  in $<<Q\Psi_0;\Psi>>=<Q\Psi_0(0){I}\circ\Psi(\infty)>$
 (in our case, $(\partial{X})^p$), only the pole at $w$ contributes to the overlap
function , so the $z$-integral's contribution to the infinitezimal overlap transformation is
\begin{eqnarray}
\sum_{j=1}^p\partial_{w}\lbrack{{\epsilon(w)}\over{(w-\xi)^{n_j+1}}}\rbrack=
\sum_{j=1}^p{{\partial\epsilon(w)}\over{(w-\xi)^{n+1}}}-(n_j+1){{\epsilon(w)}\over{(w-\xi)^{n_j+2}}}
\end{eqnarray}
This is easily integrated to give the finite transformation, i.e. the overlap function for $I(z)$:
\begin{eqnarray}
U_0(w)=\prod_{j=1}^p{{{{dI}\over{dz}}|_{z=w}}\over{(I(w)-I(\xi))^{n_j+1}}}|_{\xi=0;w\rightarrow\infty}
=w^{N-p}
\end{eqnarray}
Multiplying by the overlap function thus precisely cancels the vanishing $w^{p-N}$-factor discussed above,
keeping the correlator finite and relating the correlators before and after the conformal transformations.
The correlator is then easy to compute, as each given combination $n_1....n_p$, divided by $p!$, 
contributes exactly 1 to the correlator. Therefore the overall correlator simply equals the number 
of such combinations, i.e. the number of partitions $\lambda(N|p)$ of number $N$ with the length $p$:

\begin{eqnarray}
\sum_{N|n_1...n_p}<{{\partial^{n_1}X...\partial^{n_p}X(0)}\over{n_1!..n_p!}}(I\circ(\partial{X})^p)(\infty))>
=\lambda(N|p)
\end{eqnarray}
Next, apply the conformal transformation $f(z)=e^{iz}$ to the correlator (2.27).
Similarly to the explained above, the correlator computed on singularoid  is contributed by
the pure Schwarzian terms only with the overlap function computed to be
\begin{eqnarray}
U_0(w)={{p!w^{-2p}}\over{((p-1)!!)^2(S_{1|1}(e^{iz};z))^p}}+O(e^{iw})
\end{eqnarray}
where the Schwarzian of the exponential transformation $S_{1|1}(e^{iz};z)$ is simply ${1\over{12}}$.
Therefore the  correlator (2.27) computed on the singularoid, is given by
\begin{eqnarray}
{{\sum_{N|n_1...n_p}S_{n_1...n_p}}\over{(p-1)!!(S_{1|1}(e^{iz};z))^{p\over2}}}
\end{eqnarray}
and we deduce
\begin{eqnarray}
{\sum_{N|n_1...n_p}S_{n_1...n_p}}={\lambda(N|p)(p-1)!!(S_{1|1}(e^{iz};z))^{p\over2}}
\end{eqnarray}
 This identity particularly expresses the number of partitions of  the length $p$
in terms of summations over Stirling numbers of the second kind.
It is now straightforward to get the OSFT analytic solution of the form (2.3) for $\Psi_0$.
First of all, it is necessary to pick $\alpha_{n_1...n_p}=0$ for any $p$ odd, since
for the odd $p$ values the factorization of the OSFT correlator $<<Q\Psi_0;\Psi>>$
doesn't appear to exist. For even $p$, writing $p=2k$, the family of solutions for $\Psi_0$ is

\begin{eqnarray}
\Psi_0^{(p,q)}=c\sum_{N=2}^\infty{{\beta_{rs}(N)}\over{\lambda(N)}}\sum_{k=1}^{\lbrack{N\over2}\rbrack}
\sum_{N|n_1...n_{2k}}\prod_{j=1}^{2k}{{\partial^{(n_j)}X}\over{{n_j!\sqrt{12}}}}
\nonumber \\
\lambda(N)\equiv\sum_{k=1}^{\lbrack{N\over2}\rbrack}(2k-1)!!\lambda(N|2k)
\nonumber \\
\beta_{rs}(N)={{(N-1)^{-r}\zeta(s-1)-(N-1)^{-s}\zeta(r-1)}}
\nonumber \\
\end{eqnarray}
where $\zeta$ is the Riemann's zeta-function
and the $\lambda(N)$ coefficients are sums over the partitions of $N$ with even lengths
$2k$, weighted with $(2k-1)!!$; and $r,s$ are positive numbers greater than 2.
Indeed, it is now easy to check that, with $\Psi_0$ given by (2.31)
one has
\begin{eqnarray}
<<Q\Psi_0,\Psi>>=G_\psi\sum_{N=2}^\infty({{\zeta(s-1)}\over{(N-1)^{r-1}}}-{{\zeta(r-1)}\over{(N-1)^{s-1}}})
\nonumber \\
=G_\psi(\zeta(r-1)\zeta(s-1)-\zeta(s-1)\zeta(r-1))=0
\end{eqnarray}
This OSFT solution is straightforward to generalize to $D$ space-time dimensions;
one just has to take the product of $D$ copies  of $\Psi_0$:
\begin{eqnarray}
\Psi_0^{(D|rs)}=c\prod_{m=1}^D\Psi_0^{(m)}
\nonumber \\
\Psi_0^{(m)}=\Psi_0(X\rightarrow{X_m})
\end{eqnarray}
($X$ is replaced with $X_m$ in (2.31) with the $c$-ghost factor removed)

The summation over $N$ is essentially the summation over space-time spin values coinciding with
conformal dimensions of the string field's components; the components
 with different $k$ with $N$ fixed could then be understood as Stueckelberg terms for a given spin $N$.
Unlike the elementary BRST cohomology solutions of (1.3) defining wavefunctions of pure states, with given spins and masses,
 the solution (2.31), (2.33) sums
over the ensemble of the states with different spins and masses , with the coefficients defining the reduced density
matrix of a certain subsystem.
As our solution carries he $b-c$ ghost number $1$( and in fact can be extended to superstring theory
with no coupling to the $\beta-\gamma$  ghost system), it belongs to the same ghost sector as the generators
of Poincare isometries in space-time.
It is therefore tempting to identify  the solution (2.31), (2.33) with the reduced density matrix of the subsystem
of the lower spin 1  entangled with tower of higher spins in open string theory, with the 
terms at a given $N$ corresponding to contribution from the spin $N$ subsystem to the classical entanglement.
As the solutions (2.31), (2.33) acting on the vacuum define normalizable mixed states, it is natural to choose the normalization
so that the traces of the density matrices describing these states are equal to one $Tr(\rho)=1$.
Define the normalization factors:
\be
\lambda_{rs}^{(0)}=
\sum_{N=2}^\infty{{\beta_{rs}(N)}\over{\lambda(N)}}
\ee
These series converge fast as the partition numbers grow at least as fast as exponentially with ${\sqrt{N}}$.
The properly normalized solutions for the mixed state are
\bea
\Psi_{normalized}^{(D|rs)}=(\lambda_{rs}^{(0)})^{-D}\Psi_0^{D|rs}
\eea
This concludes  our description of the BRST-invariant mixed states appearing in open  string theory.

\section{\bf Conclusions}

In this letter we have described mixed state solutions in  linearized string field theory
(or nontrivial BRST cohomology elements defined by $Q\Psi_0=0$). Unlike the elementary solutions defining
wavefunctions of pure states in the string spectrum (such as a photon), the solutions considered in this letter
describe the operators leading  to mixed states,  given by summations over ensembles of wavefunctions with different
spins and momenta. The coefficients in the formal series, defining these solutions, correspond to the eigenvalues of the 
corresponding density matrices. In this letter, for the sake of brevity, we did not essentially
discuss physical properties of these solutions and did not address questions like related entanglement entropy
of string states, structure constants, etc. This is left for the future work, to appear soon. 
The solutions are parametrized by two numbers $r$ and $s$ entering the arguments of the $\zeta$-functions.
To understand their meaning, it is necessary to study the deformation of this class of solutions in
the interacting string field theory and, in particular,  to understand the structure of BRST cohomologies deformed by 
the solutions of the class (2.35). Some preliminary results in our work (currently in progress) suggest that these solutions
may be related to the geometrical deformations of the background from flat to dS or $AdS$ in background-independent theory,
with the cosmological constant being the function of $r$ and $s$. In particular, if the cosmological constant
is positive,this requires either $r$ or $s$ to be negative, so the $\zeta$-functions appearing in the solution
need to be regularized. We hope to address these (as well as many other related issues) in the future works
to appear.

\section{\bf Acknowledgements}

The author gratefully acknowledges the support of National Science Foundation of China (NSFC) under the project
11575119.

\end{document}